\documentclass[12pt]{article}
\input epsf      

\newcommand \no {\noindent}

\newcommand{\ba}[1]{\begin{array}{#1}}
\newcommand{\ea}{\end{array}}

\newcommand{\be}{\begin{equation}}
\newcommand{\ee}{\end{equation}}
\newcommand{\bea}{\begin{eqnarray}}
\newcommand{\eea}{\end{eqnarray}}
\newcommand{\beann}{\begin{eqnarray*}}
\newcommand{\eeann}{\end{eqnarray*}}

\def\reff#1{(\ref{#1})}
\begin{document}

\title{A faster implementation of the pivot algorithm for self-avoiding walks}

\author{Tom Kennedy
\\Department of Mathematics
\\University of Arizona
\\Tucson, AZ 85721
\\ email: tgk@math.arizona.edu
}
\maketitle

\begin{abstract}

The pivot algorithm is a Markov Chain Monte Carlo algorithm for simulating
the self-avoiding walk. At each iteration a pivot which produces a 
global change in the walk is proposed. If the resulting walk is self-avoiding,
the new walk is accepted; otherwise, it is rejected. 
Past implementations of the algorithm required a time $O(N)$ per
accepted pivot, where $N$ is the number of steps in the walk.
We show how to implement the algorithm so that the 
time required per accepted pivot is $O(N^q)$ with $q<1$. 
We estimate that $q$ is less than $0.57$ in two dimensions,
and less than $0.85$ in three dimensions. Corrections to the 
$O(N^q)$ make an accurate estimate of $q$ impossible. They also imply 
that the asymptotic behavior of $O(N^q)$ cannot be seen for walk 
lengths which can be simulated. In simulations the 
effective $q$ is around $0.7$ in two dimensions and $0.9$ in three
dimensions. Comparisons with simulations that use the standard implementation 
of the pivot algorithm using a hash table indicate that our 
implementation is faster by as much as a factor of 80 in two dimensions
and as much as a factor of 7 in three dimensions.
Our method does not require the use of a hash table and 
should also be applicable to the pivot algorithm for off-lattice models. 

\end{abstract}

\bigskip


\newpage

\section{Introduction}

The self-avoiding walk (SAW) is a simple model for polymers in dilute
solution. The interest in the model extends well beyond this application
since the model has critical exponents which exhibit universality
\cite{degennes,cj}. 
The pivot algorithm provides a fast Monte Carlo algorithm for simulating 
the model, and so it is an ideal laboratory for studying renormalization 
group predictions and, in two dimensions, conformal field theory predictions. 
This algorithm first appeared in the literature in 1969 \cite{lal}.
When Madras and Sokal did a detailed study of its efficiency
in 1988 \cite{msokal}, the number of papers on the pivot algorithm 
was still quite small. There is now a vast literature on applying the
pivot algorithm to the SAW, continuum (off-lattice) models,
SAW's with attractive interactions, self-avoiding polygons and star-branched
polymers. Even if one only considers papers which apply 
the pivot algorithm to the SAW on a lattice, 
the list of references is substantial \cite{mjhs} - \cite{ccpa}. 
Expository accounts of the pivot algorithm 
may be found in \cite{mslade,sokala}.

At each iteration in the pivot algorithm a pivot is proposed which 
keeps part of the walk fixed and pivots the rest of the walk. 
If the new walk is self-avoiding, the pivot is accepted.
Most of the proposed pivots are rejected, so   
in discussing the efficiency of the algorithm, 
it is important to distinguish between the time required per iteration
of the Markov chain and the time required per accepted pivot. 
The fraction of accepted pivots goes to zero with the length of the 
walk as $N^{-p}$ for some exponent $p>0$ which depends on the 
number of dimensions. So these two times differ by a factor of $N^p$.
For global observables such as the distance to the endpoint of the 
walk, it is believed that only a few accepted pivots are needed 
to produce an effectively independent walk. Thus the natural measure
of efficiency is to consider the time required per accepted pivot
since this should roughly measure the time required to produce 
an essentially independent sample of a global observable.

The most naive check for self intersections takes a time $O(N^2)$. 
By using a hash table, this check may be done in a time $O(N)$.
If one starts at the pivot point and works 
outwards, self intersections are typically found much faster. 
With this approach the average time required to check for 
self intersections is believed to be $O(N^{1-p})$ \cite{msokal}.
Thus the time per accepted pivot is $O(N)$, with possibly 
a logarithmic correction. The time required to carry out the pivot
is $O(N)$, and this need only be done once it has been decided 
that the pivot should be accepted.
Thus the best past implementations of the pivot algorithm have only
required a time $O(N)$ per accepted pivot.

It is sometimes said that the pivot algorithm cannot be any faster than 
this since it takes a time $O(N)$ to simply write down a walk with 
$N$ steps. Nonetheless, 
we will show how the pivot algorithm may be implemented so that the 
time required per accepted pivot is $O(N^q)$ with $q<1$. 
Significant corrections to the $O(N^q)$ behavior make an accurate 
estimate of $q$ impossible. We estimate that it is less than $0.57$
in two dimensions and less than $0.85$ in three dimensions.  
The ``effective'' value of $q$ depends strongly on the length of 
the walk, decreasing with the length of the walk. 
The values of $0.57$ and $0.85$ are roughly the effective values of 
$q$ for the longest walks that we can simulate. 

We have also performed simulations using the standard implementation of the 
pivot algorithm using a hash table. For the longest walks in two dimensions
we find that our implementation is faster by a factor of around 80, and 
in three dimensions by a factor of around 7.

In section two we explain our implementation of the pivot algorithm.
We consider the theoretical time our implementation requires as a function 
of $N$ in the third section. 
By theoretical time we mean the time that would be required by an ideal
computer with unlimited memory that was all equally fast.
We also study the real time required to run our implementation of the 
pivot algorithm and the real time required for the implementation using 
a hash table 
for walk lengths up to $1,000,000$ steps in two dimensions and up to 
$640,000$ steps in three dimensions. 

\section{The implementation}

The pivot algorithm is a Markov Chain Monte Carlo method for simulating
the SAW. It defines a Markov chain on the set of walks of a fixed length
starting at the origin so that the stationary distribution of the chain
is the uniform distribution on the walks. One can then generate samples 
of the SAW by running the Markov chain. 
An iteration of the Markov chain starts by picking a random site on the walk. 
Then one picks a random lattice symmetry $g$.
The section of the walk from the starting point to the randomly chosen 
site is not changed. The rest of the 
walk is ``pivoted'' by applying $g$ to it with respect to the randomly 
chosen site. 
This algorithm trivially satisfies detailed balance if the probabilities 
of choosing $g$ and $g^{-1}$ are equal, and it is not hard to show it is 
ergodic, i.e., the Markov chain is irreducible \cite{mslade,sokala}.

We denote the sites in the walk by $\omega(i)$ with $0 \le i \le N$. 
Our walks start at the origin, so $\omega(0)=0$. 
The nearest neighbor constraint means that 
$||\omega(i)-\omega(i-1)||=1$ for $1 \le i \le N$. 
The self-avoiding  constraint means that 
$\omega(i) \ne \omega(j)$ for $i \ne j$. 
It is misleading to think of the index $i$ as a time, so we will refer 
to such indices as ``locations'' along the walk, rather than times. 

There are two main steps in the pivot algorithm, and both 
limit its performance to $O(N)$ in current implementations. 
The first is the test for self intersections
to see if the new walk should be accepted. The second is actually 
carrying out the pivot. 

The key idea to speed up the first bottleneck is to take advantage of
the fact that the walk only takes nearest neighbor steps.
When we compare the walk at locations $i$ and $j$, we do not simply check
if $\omega(i)=\omega(j)$. Instead we compute the distance 
$d=||\omega(i)-\omega(j)||$. (The norm used should be the minimum number 
of nearest neighbor steps needed to get from $\omega(i)$ to $\omega(j)$.
On the square and simple cubic lattices this is just the $l^1$ norm.) 
If $d$ is nonzero then we can 
conclude not just that $\omega(i) \ne \omega(j)$, but also that 
\be
\omega(i^\prime) \ne \omega(j^\prime), \quad if \quad |i-i^\prime| 
+ |j-j^\prime| < d
\label{observe}
\ee
Thus we can rule out a large number of potential self intersections 
if $d$ is large.
By itself this observation is rather useless; one also needs an algorithm
for deciding which values of $i$ and $j$ to check. 
This observation with $i^\prime=i$ was used in \cite{sfg},
but the resulting algorithm did not do better than $O(N)$. 
A hierarchical algorithm for choosing the $i$ and $j$ was used in 
\cite{jmf} to obtain an efficiency of $O(N)$ per attempted pivot.

Before explaining our algorithm, we will explain the algorithm in 
\cite{sfg} since it helps illustrate how observation \reff{observe}
can be used. Let $l$ be the location at which the pivot is done. 
Fix an $i$ between $l+1$ and $N$. We need to 
check if $\omega(i)$ is equal to $\omega(j)$ for any $j$ between 
$1$ and $l-1$. The naive approach would be to simply do 
the check for all values of $j$ running from $l-1$ down to $1$. 
The idea in \cite{sfg} is that 
at a given value of $j$, we compute $d=||\omega(i)-\omega(j)||$.
If $d=0$ we have found a self intersection and we are done. 
If $d>0$, then we know that $\omega(j)$ is not equal to $\omega(i)$, 
but we also know that $\omega(j-1),\cdots,\omega(j-d+1)$ are all not
equal to $\omega(i)$ since the walk only takes nearest neighbor 
steps. So instead of decreasing $j$ by just $1$, we can decrease it 
by $d$. This will produce a dramatic speed-up compared to the naive
approach of checking all $i$ and $j$, but it will still take 
at least $O(N)$ operations to check a walk with no self-intersections
since we must consider all values  
of $i$ from $l+1$ to $N$. (Actually, the algorithm requires significantly 
more operations than that.) 

Our algorithm is as follows. 
As before, we let $l$ be the location at which the pivot is done.
Throughout the algorithm, $i$ and $j$ will be locations with $j<l<i$
which have the property that 
\be 
\omega(i') \ne \omega(j') \quad \forall \, i', j' \,
{\rm such \ that} \, j<j'<l<i'<i
\label{induct}
\ee
Initially, $i=l+1$ and $j=l-1$. 
At each step the algorithm either decreases $j$ 
or increases $i$ in such a way that property \reff{induct} remains
true. (Of course, in trying to do this we may find a self intersection, in
which case the test for self-intersections ends.)
The procedures for increasing $i$ and decreasing $j$ are completely
analogous. We will only explain the procedure for increasing $i$.
Let $m_i$ be the distance from $\omega(i)$
to $\{\omega(k): j < k < l\}$, i.e.,
\be
m_i=\min\{||\omega(i)-\omega(k)||: j < k < l\}
\ee
If $m_i=0$, then there is a self intersection. 
If $m_i>0$, then we know $\omega(i') \ne \omega(j')$ for all 
$i',j'$ with $i \le i' < i+m_i$ and $j < j' < l$. 
So instead of just increasing $i$ by $1$, we can increase it by $m_i$. 
We do not need to compute $m_i$ exactly. If we can compute a lower 
bound $b_i$ on $m_i$, then we can increase $i$ by $b_i$. 

We use a loop on $j'$ running from $l-1$ down to $j$ to 
compute a lower bound $b_i$ on $m_i$ as follows. 
At the start of the loop we set $b_i=N$. 
Before we compare $\omega(i)$ and $\omega(j')$,
$b_i$ will be a lower bound on the distance from $\omega(i)$ to 
$\{\omega(k): j'+1 \le k < l\}$. 
Let $d=||\omega(i)-\omega(j')||$.
Pick an integer $s$ with $s < d$. Then the distance from $\omega(i)$ to 
$\{\omega(k): j'-s \le k \le j'\}$ is at least $d-s$. 
So if we replace $b_i$ by $\min\{b_i,d-s\}$, then 
$b_i$ is a lower bound on the distance of $\omega(i)$ to 
$\{\omega(k): j'-s \le k < l\}$. 
So we can reduce $j'$ by $1+s$. 
When $j'$ reaches $j$, $b_i$ will be a lower bound on $m_i$. 

There are a lot of choices for how to choose the integer $s$. 
Recall that the only constraint is that $s<d$ where 
$d=||\omega(i)-\omega(j')||$. 
The simplest choice of $s$ is to take it to be $d/2$. 
(If $d$ is odd we round $d/2$ down to get $s$.)
However, we have found that the algorithm is significantly faster
with the following choice. 
If $d<b_i$ we take $s=d/2$. 
However, if $d \ge b_i$, then we take $s=d-b_i$. 
This leaves $b_i$ unchanged and reduces $j'$ by $1+d-b_i$.

Our check for self-intersections will end when $j \le -1$ and $i \ge N+1$. 
If we reach this point we know the pivoted walk should be accepted.
As long as $j > -1$ and $i < N+1$, 
we are free to choose whether we attempt to increase 
$i$ or decrease $j$. The choice we make  
is to attempt to increase $i$ if it is closer to $l$, 
and attempt to decrease $j$ otherwise.
This choice means 
that $i$ and $j$ move away from $l$ at roughly the same rate. 
So the algorithm checks for self intersections near $l$ before 
it checks for self intersections that involve locations far from $l$.
Once $j \le -1$ or $i \ge N+1$ there is no choice. We can only 
increase $i$ in the former case and decrease $j$ in the latter.

Since it takes a time $O(N)$ to simply write down a walk with $N$ steps,
it seems that the second bottleneck of carrying out an accepted 
pivot must limit the performance to $O(N)$. 
To do better, the key idea is to not 
carry out the pivot each time a pivot is accepted. Instead we
keep track of which pivots have been accepted and only carry them 
out after a certain number have been accepted. 
This implies that we do not store the present walk, but we will need to 
know some of the sites in it to determine if we should accept the next pivot. 
Thus we must store our record of the past pivots in a form which 
makes it possible to compute individual $\omega(i)$'s efficiently.

A pivot operation acting on the walk $\omega$ produces a new walk $\bar \omega$
by the equation 
\be
\bar \omega(j)= \cases { \omega(j), &for $j \le l$ \cr
          g [\omega(j)-\omega(l)] + \omega(l),  &for $j \ge l$} \label{pivotop}
\ee
Here $l$ is some location with $0 \le l < N$, which we will refer to as the 
pivot location. $g$ is a lattice symmetry which fixes the origin, i.e., 
a linear transformation which maps the lattice back into itself.
(For a given lattice, there are only a finite number of possible $g$. )
The pivot is completely determined by $l$ and $g$, so one could keep 
track of the pivots that have been accepted by simply keeping a 
list of the $l$'s and $g$'s. This is not what we 
do because the time required to compute the position of 
a location on the walk would be significant.
Instead we represent the current walk in the following way.

Suppose that we have accepted $n$ pivots and the pivot 
locations are  $l_1 < l_2 < \cdots l_n$. (Note that they are in increasing 
order, not in the order in which they were proposed and accepted.)
Let $\omega$ be the walk after these pivots, and $\omega^\prime$ the 
walk before these pivots. We can think of the segment of  $\omega^\prime$ 
from locations $l_i$ to $l_{i+1}$  as being rigid. The corresponding segment 
of $\omega$ is obtained by applying a single lattice symmetry and 
translation to the segment in $\omega^\prime$.  
This motivates representing the walk $\omega$ by the following data structure.
It consists of 

(i) the ``old'' walk $\omega^\prime$. This is the walk some number of 
iterations prior to the present. 

(ii) an integer $n$ which is the number of pivots that have been 
accepted but not carried out yet

(iii) pivot locations, $l_1 < l_2 < \cdots < l_n$ 

(iv) lattice symmetries, $g_1, g_2, \cdots, g_n$

(v) lattice sites, $x_1, x_2, \cdots, x_n$

\noindent The walk $\omega$ is obtained from this 
data structure by the equation
\be
\omega(j)=g_i \omega^\prime(j)+x_i, \quad for \quad l_i \le j \le l_{i+1}
\label{eqdata}
\ee
with $l_0$ defined to be $0$ and $l_n$ to be $N$.
There is some redundancy in this data structure. Using the fact 
that $\omega$ must be a nearest neighbor walk which starts at the 
origin, one can determine the $x_i$ from the rest of the data structure. 
However, this requires a significant amount of time, and it is faster
to simply include the $x_i$ with the data structure.

When a pivot is proposed, we do not immediately insert it in the 
lists in (iii), (iv) and (v). Let $\omega$ be the walk before the 
pivot and $\bar \omega$ the walk after the pivot. 
To test if the proposed pivot should be accepted we need to be 
able to compute $\bar \omega(j)$ for selected 
values of $j$. We use the data structure and \reff{eqdata} 
to compute $\omega(j)$ and then use \reff{pivotop} to compute 
$\bar \omega(j)$. 

Now suppose that we have accepted a pivot with pivot location $l$ and 
lattice symmetry $g$. 
So the new walk $\bar \omega$ is given by \reff{pivotop}.
We must determine how to update the data structure. 
We do this in two steps. 
First we simply add the pivot location to the list. 
Let $k$ be such that $l_k < l < l_{k+1}$. Then the changes in the data 
structure are 

(ii) $n \rightarrow n+1$

(iii) $l_1, l_2, \cdots, l_n \rightarrow 
l_1, l_2, \cdots, l_k, l, l_{k+1}, \cdots, l_n $ 

(iv) $g_1, g_2, \cdots, g_n \rightarrow 
g_1, g_2, \cdots, g_{k-1}, g_k, g_k, g_{k+1}, \cdots, g_n $

(v) $x_1, x_2, \cdots, x_n \rightarrow
x_1, x_2, \cdots, x_{k-1}, x_k, x_k, x_{k+1}, \cdots, x_n $

\noindent At this stage the data structure still represents the walk 
before the pivot. It is possible that $l$ is equal to one of 
the $l_i$. In this case we simply skip step one. 

The second step is to carry out a pivot under the assumption that the 
pivot location $l$ is in the list $l_1,l_2, \cdots l_n$. 
Let $l=l_k$. 
Suppose that $j$ satisfies $l_i \le j \le l_{i+1}$ with $i \ge k$. 
Then
\be
\bar \omega(j) 
= g[\omega(j)-x]+x = g[g_i \omega^\prime(j) + x_i - x]+x
= g g_i \omega^\prime(j) + g x_i - g x + x
\ee
where $x=\omega(l_k)$. 
Thus for $i \ge k$, 
\be 
g_i \rightarrow g g_i
\label{gupdate}
\ee
\be
 x_i \rightarrow g x_i - g x + x 
\label{xupdate}
\ee
For $i<k$, $g_i$ and $x_i$ are left unchanged.
In both steps the ``old'' walk $\omega^\prime$ is not changed. 

We use this data structure to carry out the pivot algorithm as 
follows. We fix a large integer $N_{pivot}$ which will be small 
compared to $N$, the number of steps in the walk. 
As pivots are accepted, we update the data structure as explained above. 
When $n$ reaches $N_{pivot}$, we use \reff{eqdata} to compute 
the walk $\omega$ and replace $\omega^\prime$ in the data structure 
with $\omega$. Then we set $n=0$ and delete the lists in 
(iii),(iv), and (v). Note that the integer $n$ in (ii) of the data 
structure is not always equal to the number of pivots that have 
been accepted but not yet carried out. It can be slightly less 
since the pivot location will sometimes already be present 
in the list in (iii).

As we said before, this data structure is only useful if we can 
compute $\omega(j)$ quickly for a given $j$. 
This is done with equation \reff{eqdata}.
The nontrivial part is finding the $i$ so that $l_i \le j \le l_{i+1}$. 
This can be done in a time of order $\ln(n)$. 

\section{Analysis}

There are three steps in the pivot algorithm for which the time required
depends on the length of the walk.

\smallskip

\no 1. For each proposed pivot we must decide whether to accept the pivot 
or not.

\smallskip

\no 2. For each accepted pivot we must update elements (ii) to (v) 
in the above data structure. 

\smallskip

\no 3. For every $N_{pivot}$ accepted pivots we must carry
out the pivots implicit in the above data structure, i.e., we must update
element (i) in the data structure.

\medskip

One cannot determine a priori how many steps will be needed in the test for
self-intersections, so the analysis of the first
step will require some empirical study.  
In this step we must use \reff{eqdata} repeatedly. In this equation we 
are given $j$ and must find $i$ so that $l_i \le j \le l_{i+1}$
The lists in (iii), (iv) and (v)
are stored as linear arrays in the order given by the condition
$l_1 < l_2 < \cdots < l_n$. 
Using a bisection procedure, 
$i$ may be found in a time of order $\ln(n)$, which on average is 
$O(\ln(N_{pivot}))$. 
Let $D(N)$ denote the number of times the distance 
$||\omega(i)-\omega(j)||$ is computed {\it per accepted pivot}. 
We assume that $D(N)$ grows as $N^\sigma$. 
Except for the steps involved in finding $i$ in \reff{eqdata}, 
the number of steps needed to check for self-intersections per accepted 
pivot is proportional to $D(N)$.  
So the time required for the first step will be $O(N^\sigma \ln(N_{pivot}))$
per accepted pivot.
We will estimate $\sigma$ by running simulations,
counting the number of times $||\omega(i)-\omega(j)||$ must be 
computed, and dividing this number by the number of accepted pivots.

Now consider the second step. 
Since the lists in (iii), (iv) and (v)
are stored as linear arrays, the number of 
operations required to insert the new entries is of order $n$.
We must also carry out the updates given by \reff{gupdate} and 
\reff{xupdate}. On average this takes order $n$ operations too. 
The insertion of the new entries could be done more quickly with a more 
sophisticated data structure, but since \reff{gupdate} and 
\reff{xupdate} take order $n$ operations, it is not clear that 
the improvement would be significant.
So the time required for the second step is of order $n$,  
and since $n$ increases from $0$ to $N_{pivot}$, 
on average the time per accepted pivot for this step is $O(N_{pivot})$. 

Finally, we consider the third step. 
The walk in (i) is updated using \reff{eqdata}.
The search for the index $i$ can be avoided here. 
We can simply loop on $i$ and then 
within the loop on $i$ we loop on $j=l_i,\cdots,l_{i+1}$ .
Thus each application of \reff{eqdata} only takes a time of 
order 1 and so the update of the walk takes a time of order $N$. 
Since this need only be done 
for every $N_{pivot}$ accepted pivots, the time per accepted pivot is
$O(N/N_{pivot})$. 

Thus the total time per accepted pivot is 
\be 
O(N^{\sigma} \ln(N_{pivot})) + O(N_{pivot}) + O({N \over N_{pivot}}) 
\ee
We take $N_{pivot}$ to be proportional to $\sqrt{N}$. 
In our simulations we have found that $N_{pivot}=\sqrt{N/40}$ 
is a good choice for $N_{pivot}$, both in two and three dimensions. 
(There is a fairly wide range of values of $N_{pivot}$ for which the 
algorithm takes roughly the same amount of time.)
Then the total time per accepted pivot is 
\be 
O(N^{\sigma} \ln(N)) + O(N^{1/2})
\ee

We now turn to the simulations to estimate $\sigma$.
We start our simulations with a walk that is a straight 
line. This is a very atypical walk. For example, the fraction of accepted 
pivots is significantly higher when one starts with this walk until
the walk is well ``thermalized.'' So it is important to first run 
the algorithm until this initialization bias is removed. 
We do this as follows. We compute the fraction of locations in the walk
at which the walk turns rather than goes straight. This fraction starts at 0.
This fraction is a random variable, but for long walks its variance is 
quite small.  
We run the algorithm until this fraction has reached roughly $90\%$
of its equilibrium value. Then we run the algorithm for a total of 
ten times the number of iterations required to reach $90\%$. 
This is then the starting point for our estimation of $\sigma$. 

\begin{table}[thp]
   \begin{center}
      \begin{tabular}{ || r | r | r | r ||}
         \hline
         N \, \, \, & $\ln(D(N))$ \, \, \,  & Acceptance fraction
          \\ \hline\hline
     1,000 & 5.861836 $\pm$  0.000175   &  0.253622 $\pm$  0.000037 \\ \hline 
     1,600 & 6.180086  $\pm$ 0.000189   &  0.231511 $\pm$  0.000035 \\ \hline 
     2,500 & 6.476449  $\pm$ 0.000213   &  0.212222 $\pm$  0.000033 \\ \hline 
     4,000 & 6.782237  $\pm$ 0.000232   &  0.193764 $\pm$  0.000034 \\ \hline 
     6,400 & 7.082294  $\pm$ 0.000248   &  0.176967 $\pm$  0.000032 \\ \hline 
    10,000 & 7.362418 $\pm$  0.000251   &  0.162461 $\pm$  0.000031 \\ \hline 
    16,000 & 7.653438  $\pm$ 0.000279   &  0.148499 $\pm$  0.000031 \\ \hline 
    25,000 & 7.925862  $\pm$ 0.000309   &  0.136442 $\pm$  0.000031 \\ \hline 
    40,000 & 8.209963  $\pm$ 0.000314   &  0.124717 $\pm$  0.000028 \\ \hline 
    64,000 & 8.489480  $\pm$ 0.000345   &  0.114123 $\pm$  0.000028 \\ \hline 
   100,000 & 8.753149 $\pm$  0.000382   &  0.104931 $\pm$  0.000029 \\ \hline 
   160,000 & 9.027825  $\pm$ 0.000371   &  0.095989 $\pm$  0.000028 \\ \hline 
   250,000 & 9.286426  $\pm$ 0.000408   &  0.088225 $\pm$  0.000026 \\ \hline 
   400,000 & 9.555579  $\pm$ 0.000432   &  0.080820 $\pm$  0.000024 \\ \hline 
   640,000 & 9.822300  $\pm$ 0.000455   &  0.074010 $\pm$  0.000025 \\ \hline 
 1,000,000 & 10.074441 $\pm$ 0.000450   &  0.068058 $\pm$  0.000023 \\ \hline 
      \end{tabular}
      \caption{\protect Square lattice: $D(N)$ is the number of 
        distance computations required per accepted pivot. 
        The acceptance fraction is the ratio of the number of accepted 
        pivots to the number of attempted pivots. 
        (Error bars are one standard deviation.)
        }
      \label{table_sq_d}
   \end{center}
\end{table}

\begin{table}[thp]
   \begin{center}  
      \begin{tabular}{ || r | r | r ||}
         \hline
         N \, \, & $\ln(D(N))$ \, \, \,  & Acceptance fraction
          \\ \hline\hline
     1,000 & 6.557023 $\pm$ 0.000269   &  0.461623 $\pm$  0.000092  \\ \hline
     1,600 & 6.987372 $\pm$ 0.000386   &  0.437485 $\pm$  0.000105  \\ \hline
     2,500 & 7.391570 $\pm$ 0.000451   &  0.415891 $\pm$  0.000108  \\ \hline
     4,000 & 7.813830 $\pm$ 0.000427   &  0.394216 $\pm$  0.000082  \\ \hline
     6,400 & 8.232902 $\pm$ 0.000396   &  0.373658 $\pm$  0.000103  \\ \hline
    10,000 & 8.627014 $\pm$ 0.000468   &  0.355183 $\pm$  0.000104  \\ \hline
    16,000 & 9.039961 $\pm$ 0.000453   &  0.336693 $\pm$  0.000100  \\ \hline
    25,000 & 9.428706 $\pm$ 0.000473   &  0.320281 $\pm$  0.000090  \\ \hline
    40,000 & 9.836744 $\pm$ 0.000551   &  0.303548 $\pm$  0.000104  \\ \hline
    64,000 & 10.242274 $\pm$ 0.000557  &  0.287784 $\pm$  0.000110  \\ \hline
   100,000 & 10.625056 $\pm$ 0.000654  &  0.273644 $\pm$  0.000102  \\ \hline
   160,000 & 11.026357 $\pm$ 0.000710  &  0.259694 $\pm$  0.000122  \\ \hline
   250,000 & 11.406323 $\pm$ 0.000605  &  0.246912 $\pm$  0.000101  \\ \hline
   400,000 & 11.804262 $\pm$ 0.000698  &  0.234145 $\pm$  0.000092  \\ \hline
   640,000 & 12.200080 $\pm$ 0.000693  &  0.222179 $\pm$  0.000099  \\ \hline
      \end{tabular}
      \caption{\protect Simple cubic lattice: the same quantities 
        are shown as in table \ref{table_sq_d}.
        }
      \label{table_cubic_d}
   \end{center}   
\end{table} 

In the two tables we give the results of simulations 
with our implementation of the pivot algorithm for walk lengths ranging from 
1,000 steps to 1,000,000 steps on the square lattice
and 1,000 steps to 640,000 steps on the simple cubic lattice. 
These results for the square lattice are based on 271 million iterations
after the thermalization;
for the simple cubic lattice they are based on 52 million iterations
after the thermalization.
The quantity $D(N)$ is the number of distance computations that 
must be done per accepted pivot. 
The quantity shown in the tables is the logarithm of $D(N)$.

The acceptance fractions for the pivot algorithm for the various values of 
$N$ are also shown in the table. 
For the simple cubic lattice our values of the acceptance fraction
are different from those in \cite{lms}. 
This is because their computations included the trivial pivot (which 
is always accepted) as a possible pivot \cite{pc}. 
If $f$ is the acceptance fraction given in their table, then 
$(48 f-1)/47$ is their acceptance fraction without the trivial 
pivot included. With this correction their values agree well with ours. 
It is important to note that while including the trivial pivot as a possible
pivot is slightly inefficient, it is a completely legitimate implementation
of the pivot algorithm. No quantities in \cite{lms} need to be corrected
other than these acceptance fractions.
All the error bars in the table are one standard deviation.

It is surprising how small $D(N)$ is, especially in two dimensions. 
For example, to check for self intersections in two dimensions, 
it takes on average only about 25,000 distance computations {\it per
accepted pivot} for walks of length 1,000,000. 
In three dimensions, the number of distance computations per accepted 
pivot for walks of length 640,000 is only about 200,000.
We believe that the reason $D(N)$ is so much smaller for two dimensions 
than for three is that the SAW is more spread out in two dimensions.

\begin{figure}[htbp]
  \begin{center}
    \leavevmode
    \epsfxsize=5in 
    \epsfbox{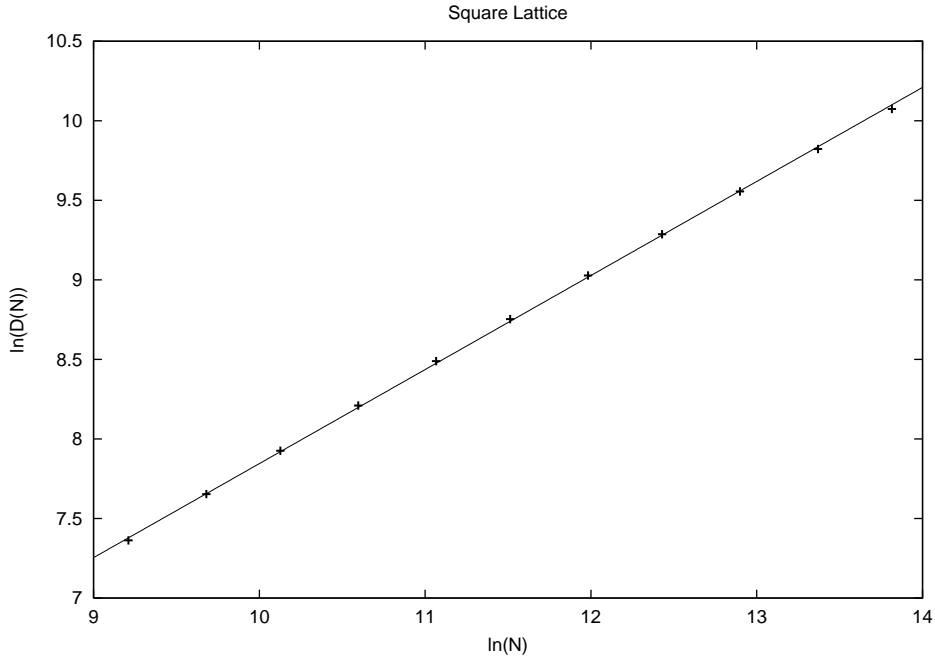}
  \end{center}
  \caption{Square lattice: $D(N)$ is the number of times we must compute 
     $||\omega(i)-\omega(j)||$ per accepted pivot.
     We plot $\ln(D(N))$ as a function of $\ln(N)$. 
     The solid line is a weighted least squares fit assuming $D(N)$ 
     is proportional to $N^\sigma$.
   }
  \label{figcomparesq}
\end{figure}

\begin{figure}[htbp]
  \begin{center}
    \leavevmode
    \epsfxsize=5in 
    \epsfbox{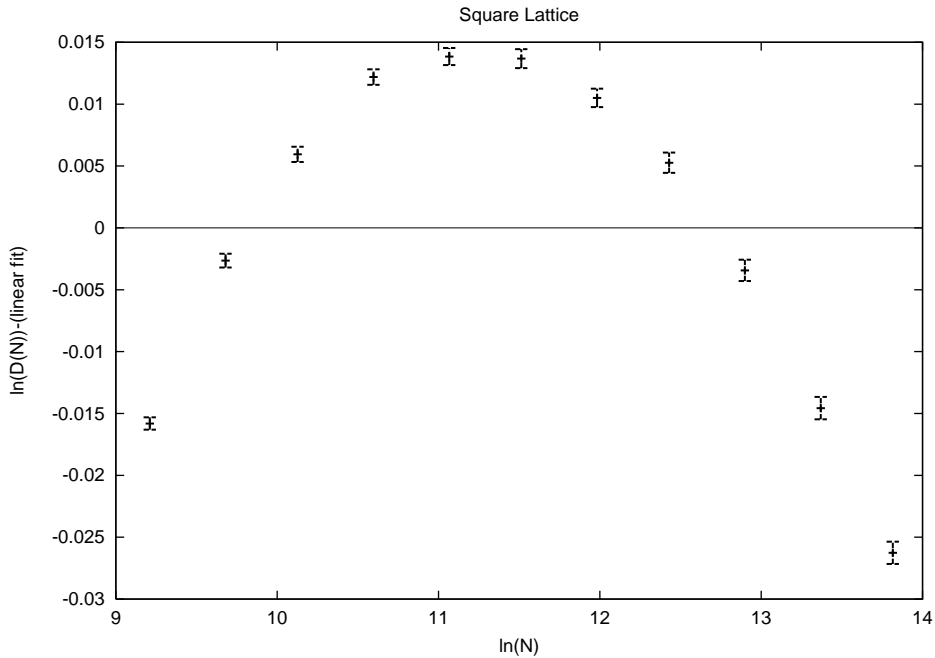}
  \end{center}
  \caption{Square lattice: The points 
    are the difference between $\ln(D(N))$ and the fit in figure 
    \ref{figcomparesq}. So the horizontal axis is the fit in figure 
    \ref{figcomparesq}. 
   }
  \label{figcomparedifsq}
\end{figure}

If $D(N)=c N^\sigma$, then
\be 
\ln(D(N))=\ln(c)+\sigma \ln(N) \label{fitlinear}
\ee
So $\sigma$ and $c$ can be found by performing a weighted least
squares fit. 
In figure \ref{figcomparesq} we show the data and the resulting 
linear fit for $N=10,000$ to $N=1,000,000$.
The data points show a systematic error with respect to the fit 
\reff{fitlinear}.
The weighted residual sum of squares (RSS) is $14292$.
If the fit were correct and the errors were independent and 
normally distributed then 
RSS would have a chi-squared distribution, and the probability of a 
value of RSS greater than $21.67$ would be $0.01$.
The systematic error is more apparent in figure \ref{figcomparedifsq} in
which we plot the difference of the data and the fit, i.e, 
$\ln(D(N))-\ln(c)-\sigma \ln(N)$.
So the fit \reff{fitlinear} is the horizontal axis in this figure.

\begin{figure}[htbp]
  \begin{center}
    \leavevmode
    \epsfxsize=5in 
    \epsfbox{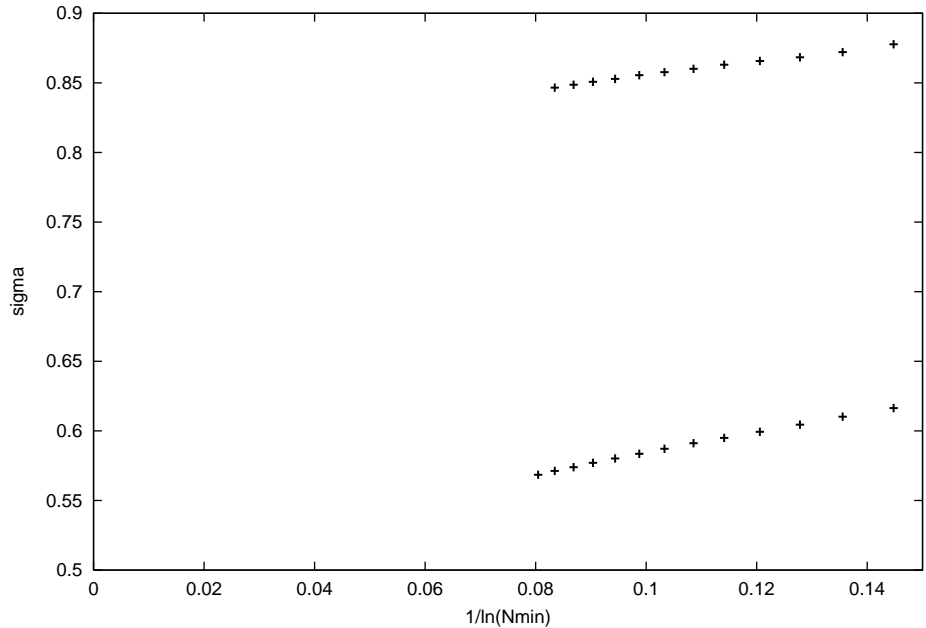}
  \end{center}
  \caption{The value of $\sigma$ that results from a weighted 
     least squares fit to \reff{fitlinear} is plotted as a function
     of $1/\ln(N_{min})$. $N_{min}$ is the length of 
     the shortest walk used in the fit.
     The bottom curve is for the square lattice; the top curve is for the 
     simple cubic lattice.
   }
  \label{figsigma}
\end{figure}

The large value of RSS found above indicates that there are significant
corrections to \reff{fitlinear}. This can also be seen by studying
how $\sigma$ changes as we vary the smallest $N$ value that is included 
in the fit. 
The bottom curve in figure \ref{figsigma} 
shows the value of $\sigma$ that comes from a 
weighted least squares fit to \reff{fitlinear} as a 
function of $1/\ln(N_{min})$ where $N_{min}$ is the smallest value of 
$N$ included in the fit.
The figure shows that our estimate of $\sigma$ depends strongly on 
$N_{min}$ and shows no sign of stabilizing for the values of $N$ that we 
can simulate. 
The RSS for all these fits are much larger than the values that would
be expected if \reff{fitlinear} was a true fit to the data. 
Because of the small statistical errors in our estimates of $D(N)$, 
the error bars for $\sigma$ that come from this fit are tiny 
compared to the change in $\sigma$ as $N_{min}$ is changed. 
These error bars are not shown in the figure. 

\newpage

One can try to improve the fit by including additional terms in 
\reff{fitlinear}. However, because of the lack of any a priori knowledge 
of the form of these corrections and the size of the corrections, 
it does not appear that a reliable estimate of $\sigma$ is possible. 
As an example, suppose we assume that 
\be
D(N)=c N^\sigma [1+ b N^{-\Delta}]
\label{fitadditivepre}
\ee
Then $\ln(D(N))$ is approximately 
\be
\ln(D(N))=\ln(c) + \sigma \ln(N) + b N^{-\Delta}
\label{fitadditive}
\ee
For a given value of $\Delta$ we can find $\sigma$ and $b$ by a weighted
least squares fit. We then search over $\Delta$ to find the value that 
gives the smallest RSS. 
However, a rather wide range of values of $\Delta$ will give an 
acceptable value of RSS and the resulting values of $\sigma$ vary 
considerably. For example, if we use $N_{min}=10,000$ 
then RSS is less than the $99\%$ confidence level for 
$\Delta$ ranging from $0.072$ to $0.240$. The resulting values of $\sigma$ 
range from $0.403$ and $0.534$. We should also note that these values
of $\Delta$ are rather small and the values of $b$ that we obtain 
are all greater than $3$.
So \reff{fitadditive} is not a good approximation to \reff{fitadditivepre}.

\newpage

\begin{figure}[htbp]
  \begin{center}
    \leavevmode
    \epsfxsize=5in 
    \epsfbox{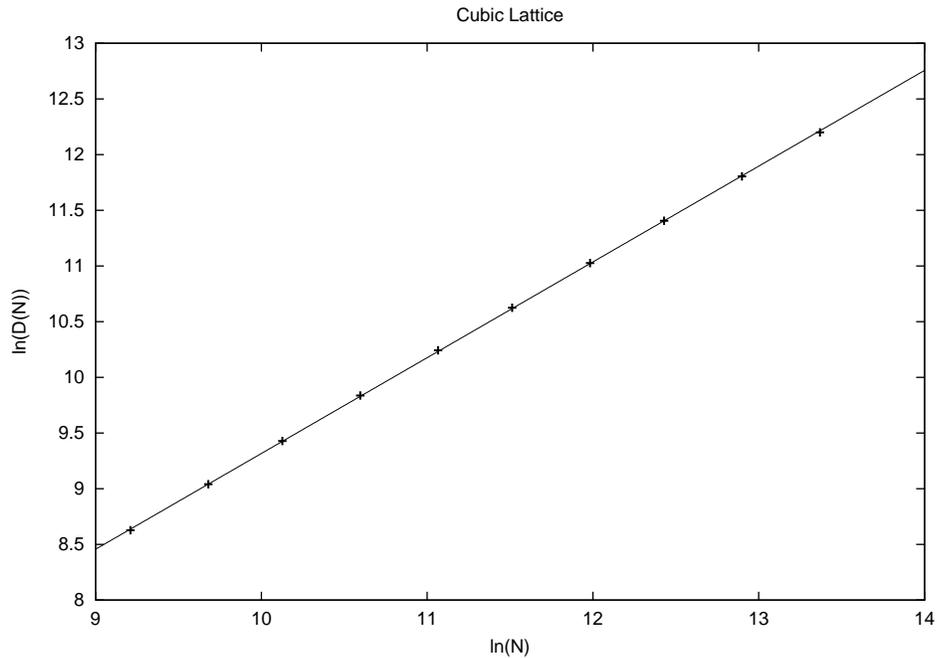}
  \end{center}
  \caption{$D(N)$ for the simple cubic lattice.
     The line is a weighted least squares fit assuming $D(N)$ is 
     proportional to $N^\sigma$.
   }
  \label{figcomparecubic}
\end{figure}

\begin{figure}[htbp]
  \begin{center}
    \leavevmode
    \epsfxsize=5in 
    \epsfbox{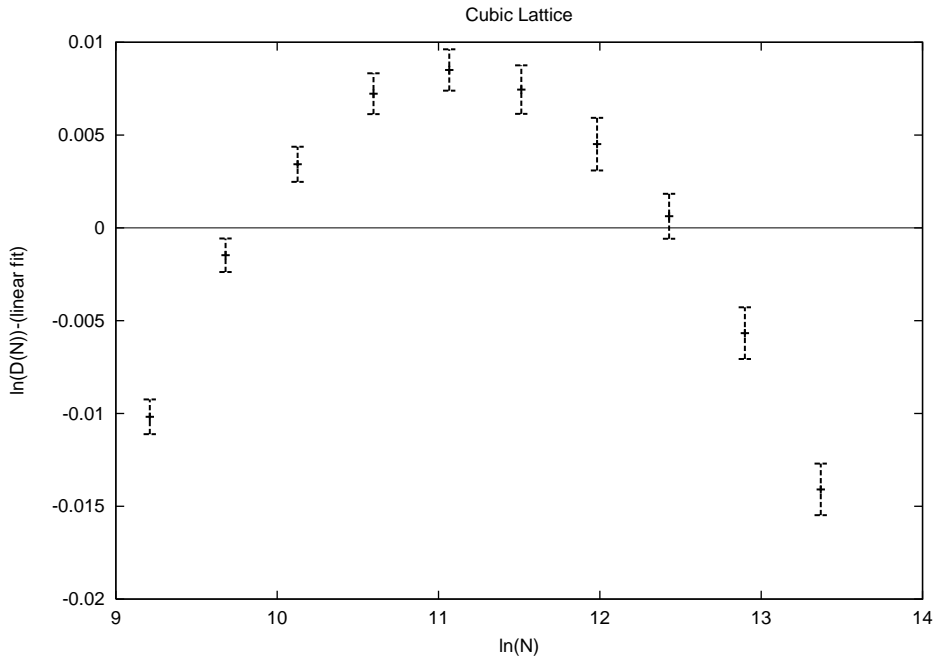}
  \end{center}
  \caption{Simple cubic lattice: The points
    are the difference between $\ln(D(N))$ and the fit in figure 
    \ref{figcomparecubic}. 
   }
  \label{figcomparedifcubic}
\end{figure}

For the simple cubic lattice the data for $N=10,000$ to $N=640,000$ 
and the fit using \reff{fitlinear} are shown in 
figure \ref{figcomparecubic}.
There is again a   
systematic error in this fit; RSS is $1591$ while the $99\%$ confidence
level is $20.09$. 
It is more clearly seen in figure \ref{figcomparedifcubic} in which  
the difference of $\ln(D(N))$ and the linear fit is plotted.
The top curve in figure \ref{figsigma} shows the value of $\sigma$ 
we obtain from the fit to \reff{fitlinear} using different values of 
$N_{min}$. As with the square lattice the value of $\sigma$
depends significantly on $N_{min}$ and shows no sign of stabilizing.

\newpage

Attempts to fit the data for the simple cubic lattice 
with \reff{fitadditive} produce meaningless results. 
The value of RSS is within the $99\%$ confidence interval for $\Delta$
ranging from $0.43$ down to almost $0$. Again, the values of $b$ we 
find are such that \reff{fitadditive} is not a good approximation to 
\reff{fitadditivepre}.

It appears impossible to reliably estimate $\sigma$. However, since
$\ln(D(N))$ appears to be a concave function of $\ln(N)$, we can give
upper bounds on $\sigma$. For two dimensions the linear 
fit \reff{fitlinear} with $N_{min}=250,000$ gives $\sigma=0.57$. 
For three dimensions, using $N_{min}=160,000$ we find $\sigma=0.85$. 
These are the values we have taken as upper bounds on $\sigma$. 
They are conservative estimates, and the true value of $\sigma$ is 
probably significantly less. 

\newpage

Our analysis has only considered the time required to produce a 
sequence of walks. In addition, one must also compute the value of 
the observable on these walks. For some observables, e.g., the 
end to end distance, this only takes a time of order 1. But for other 
observables the time required could grow with $N$ faster than the 
time needed to produce the walks. In this case it might be 
best to only compute the observable after a fixed number of Monte 
Carlo iterations. Given estimates of the autocorrelation time of the 
observable, the time needed to produce an accepted pivot,
and the time needed to compute the observable, one could determine 
the frequency for computing the observable that would give the smallest 
error bars for a given amount of CPU time. 

The previous analysis only tells us how 
the time required to run the algorithm will grow with $N$ for very large $N$. 
From a practical point of view one wants to know how much time 
our implementation of the algorithm takes and how this time compares 
with the implementation of the algorithm using a hash table to check
for self intersections.
Tables \ref{table_sq_time} to \ref{table_cubic_time_altar} compare the 
times needed to run one million iterations of the pivot algorithm
for three different implementations. 
(By one million iterations we mean one million attempted pivots, 
not one million accepted pivots.) 
One implementation is the implementation given in this paper. 
Another implementation is the implementation using a hash 
table to check for self intersections. The third implementation 
uses the test for self-avoidance given in this paper, 
but does not use the data structure (i)-(v). 
This implementation is expected to take a time $O(N)$ per 
accepted pivot. 

The column labelled ``Time'' gives the time (in secs) for the implementation 
in this paper. For the other two implementations, we give 
the ratio of the time they take to the time required by the algorithm 
of this paper.
The column labelled ``Skip'' is the implementation which uses the test for
self-avoidance given in this paper, but not the data structure
(i)-(v). The standard implementation which uses a hash table 
is shown in the columns labelled ``Hash 1'' and ``Hash 2.''
The difference between the two columns is only in the hash function used.
Letting $(x,y)$ or $(x,y,z)$ be the coordinates of the point, Hash 1 uses 
a hash table with $L=5,000,001$ entries and the hash function
\be
(x,y) \rightarrow  |(17*x+290*y)| \, mod \, L
\ee
\be
(x,y,z) \rightarrow  |(17*x+290*y+4914*z)| \, mod \, L
\ee
Hash 2 uses a hash table with $L=5,000,000$ entries and the hash function
\be
(x,y) \rightarrow  |(47*x+2210*y)| \, mod \, L
\ee
\be
(x,y,z) \rightarrow  |(47*x+2210*y+103824*z)| \, mod \, L
\ee

\begin{table}[thp]
   \begin{center}
      \begin{tabular}{ || r || r || r | r | r | r ||}
         \hline
         N \, \, & Time   &  Hash 1      & Hash 2      & Skip        & Therm. \\ \hline\hline
        1,000 & 08.61 s &   3.505  &  4.699  &  0.824        &  1  \\ \hline 
        1,600 & 09.88 s &   5.030  &  6.473  &  0.869        &  1  \\ \hline 
        2,500 & 11.26 s &   6.970  &  8.821  &  0.932        &  1  \\ \hline 
        4,000 & 13.04 s &  10.669  & 14.529  &  1.025        &  1  \\ \hline 
        6,400 & 15.11 s &  19.931  & 26.437  &  1.163        &  1  \\ \hline   
       10,000 & 18.06 s &  28.326  & 34.989  &  1.241        &  1  \\ \hline 
       16,000 & 21.39 s &  39.081  & 47.297  &  1.530        &  2  \\ \hline 
       25,000 & 25.41 s &  46.345  & 55.366  &  1.671        &  3  \\ \hline 
       40,000 & 34.45 s &  52.599  & 61.900  &  2.242        &  5  \\ \hline 
       64,000 & 53.80 s &  49.360  & 57.759  &  2.869        &  9  \\ \hline 
      100,000 & 74.26 s &  52.849  & 61.489  &  3.342        & 15  \\ \hline 
      160,000 & 102.62 s &  56.715  & 65.522   & 3.630        & 26  \\ \hline 
      250,000 & 131.82 s &  63.577  & 72.756  &  4.103        & 44  \\ \hline 
      400,000 & 169.86 s &  76.645  & 85.135  &  4.634        & 75  \\ \hline 
      640,000 & 214.35 s &  95.470  & 97.764  &  5.194        & 129  \\ \hline 
    1,000,000 & 272.33 s & 292.659  & 112.975 &   5.954       & 224  \\ \hline 
      \end{tabular}
      \caption{\protect Square lattice: 
        The column labelled ``Time'' gives the time in secs required for 
        1,000,000 iterations of our implementation of the pivot algorithm 
        on a PC running a 1.33GHz AMD processor. 
        The columns ``Hash 1'' and ``Hash 2'' 
        are for the implementation of the pivot algorithm using 
        a hash table. The two columns use different hash functions. 
        In these columns we give the ratio of the time required to 
        the time shown in the column ``Time.'' 
        The column ``Skip'' uses the test for self-avoidance given in 
        this paper, but does not use the data structure given in (i)-(v).
        It also gives the ratio of the time to the time in column ``Time.''
        The final column gives the number of iterations (in millions) 
        carried out to thermalize the walk.
        }
      \label{table_sq_time}
   \end{center}
\end{table}

\begin{table}[thp]
   \begin{center}
      \begin{tabular}{ || r || r || r | r | r | r ||}
         \hline
         N \, \, & Time   &  Hash 1      & Hash 2      & Skip \\ \hline\hline
        1,000 & 32.82 s &     2.456 &   2.753  &  1.280   \\ \hline 
        1,600 & 37.34 s &     3.216 &   3.658  &  1.372   \\ \hline 
        2,500 & 43.76 s &     4.087 &   4.692  &  1.480   \\ \hline 
        4,000 & 51.40 s &     5.352 &   6.164  &  1.623   \\ \hline 
        6,400 & 59.62 s &     7.083 &   8.149  &  1.785   \\ \hline   
       10,000 & 68.41 s &     9.594 &  10.908  &  1.986   \\ \hline 
       16,000 & 81.29 s &    13.680 &  15.621  &  2.197   \\ \hline 
       25,000 & 94.39 s &    17.614 &  19.815  &  2.499   \\ \hline 
       40,000 & 111.25 s &    23.520 &  26.075 &   2.763  \\ \hline 
       64,000 & 139.73 s &    27.786 &  30.547 &   3.174  \\ \hline 
      100,000 & 175.69 s &    32.742 &  35.819 &   3.666  \\ \hline 
      160,000 & 227.57 s &    37.587 &  40.850 &   4.333  \\ \hline 
      250,000 & 279.65 s &    44.607 &  47.504 &   4.919  \\ \hline 
      400,000 & 346.40 s &    56.945 &  58.078 &   5.719  \\ \hline 
      640,000 & 420.47 s &    79.371 &  69.534 &   6.735  \\ \hline 
    1,000,000 & 512.97 s &   378.945 &  84.269 &   7.838  \\ \hline 
      \end{tabular}
      \caption{\protect Square lattice: 
        The same quantities are shown as in the previous table. The only 
        difference is that these timing tests were done on a PC 
        running a 450 MHz Pentium II. 
        }
      \label{table_sq_time_altar}
   \end{center}
\end{table}

\begin{table}[thp]
   \begin{center}
      \begin{tabular}{ || r || r || r | r | r | r ||}
         \hline
         N \, \, & Time   &  Hash 1      & Hash 2      & Skip        & Therm. \\ \hline\hline
   1,000 &   53.7 s &   2.334 &   2.661 &   1.183      &  1  \\ \hline 
   1,600 &   70.7 s &   2.785 &   3.140 &   1.296      &  1  \\ \hline 
   2,500 &   93.4 s &   3.422 &   3.961 &   1.397      &  1  \\ \hline 
   4,000 &  128.6 s &   4.559 &   5.672 &   1.434      &  1  \\ \hline 
   6,400 &  165.1 s &   6.614 &   7.682 &   1.610      &  1  \\ \hline   
  10,000 &  234.7 s &   7.376 &   8.495 &   1.672      &  1  \\ \hline 
  16,000 &  315.7 s &   8.486 &   9.687 &   1.801      &  1  \\ \hline 
  25,000 &  451.3 s &   8.613 &   9.783 &   1.852      &  2  \\ \hline 
  40,000 &  693.6 s &   8.379 &   9.457 &   1.868      &  3  \\ \hline 
  64,000 &  1092.1 s &   8.746 &   9.801 &   1.936      &  5  \\ \hline 
 100,000 &  1605.5 s &   8.938 &   9.969 &   1.964      &  8  \\ \hline 
 160,000 &  2531.7 s &   8.640 &   9.635 &   1.953      &  12  \\ \hline 
 250,000 &  3767.4 s &   8.331 &   9.254 &   1.925      &  20 \\ \hline 
 400,000 &  5952.7 s &   8.482 &   9.415 &   1.925      &  34 \\ \hline 
 640,000 &  9857.1 s &   7.945 &   8.799 &   1.873      &  56 \\ \hline 
      \end{tabular}
      \caption{\protect Simple cubic lattice: 
        The same quantities are shown as in the previous tables.
        This table is for the computer using an AMD 1.33 GHz processor.
        }
      \label{table_cubic_time}
   \end{center}
\end{table} 

\begin{table}[thp]
   \begin{center}
      \begin{tabular}{ || r || r || r | r | r | r ||}
         \hline
         N \, \, & Time   &  Hash 1      & Hash 2      & Skip \\ \hline\hline
   1,000 & 108.5 s   &   1.956  &  2.151  &  1.378     \\ \hline 
   1,600 & 148.9 s   &   2.218  &  2.420  &  1.482     \\ \hline 
   2,500 & 196.3 s   &   2.522  &  2.760  &  1.583     \\ \hline 
   4,000 & 264.8 s   &   2.846  &  3.099  &  1.697     \\ \hline 
   6,400 & 355.2 s   &   3.406  &  3.758  &  1.804     \\ \hline   
  10,000 & 479.8 s   &   4.074  &  4.473  &  1.900     \\ \hline 
  16,000 & 638.8 s   &   5.086  &  5.517  &  1.990     \\ \hline 
  25,000 & 937.7 s   &   5.730  &  6.150  &  2.129     \\ \hline 
  40,000 & 1305.0 s  &    6.289 &   6.682 &  2.187     \\ \hline 
  64,000 & 1932.4 s  &    6.368 &   6.739 &  2.237     \\ \hline 
 100,000 & 2929.0 s  &    6.267 &   6.616 &  2.256     \\ \hline 
 160,000 & 4180.2 s  &    7.012 &   7.405 &  2.333     \\ \hline 
 250,000 & 5994.4 s  &    7.167 &   7.543 &  2.346     \\ \hline 
 400,000 & 9068.9 s  &    7.069 &   7.446 &  2.352     \\ \hline 
 640,000 & 14025.0 s &    7.099 &   7.457 &  2.325     \\ \hline 
      \end{tabular}
      \caption{\protect Simple cubic lattice: 
        The same quantities are shown as in the previous tables.
        This table is for the computer using a Pentium II 450 MHz processor.
        }
      \label{table_cubic_time_altar}
   \end{center}
\end{table} 

The tables also give the number of iterations (in millions) we ran to achieve 
thermalization. These numbers are included to give the reader an idea of
how much time is required to equilibriate walks of various lengths. 
For the longest walks the time needed for thermalization can dominate 
the total time used in the simulation.

All of the programs used in the 
timing test were written in C++, compiled with gnu compilers, and run  
under the Linux operating system. As much as possible we used the same 
data structures and programming strategies in the three programs. 
In particular, all the code is written so that lattices other than the 
square and simple cubic lattices are easily implemented. 
Walks are stored as arrays of ``points,'' where point is a class, the 
exact nature of which depends on the lattice. Lattice dependent operations
(like the lattice symmetries) are separated out in functions. 
It is natural to worry that this generality slows down the code significantly. 
To test this, we have also run timing tests with a Fortran program based 
closely on code used in \cite{lms}. This program is specifically 
written for the simple cubic lattice and uses a hash table. 
We have found that the Fortran program runs only about $5\%$ 
faster than our program using a hash table.

The time required depends of course on the type of computer used. One might 
expect that the ratios of times for the different algorithms 
do not depend much on the computer. 
However, we have found that when exactly the same 
code is run on different PC's these ratios can vary significantly.
Thus we have shown the results of timing tests on two computers.
Both are PC's. One uses an AMD 1.33 GHz processor, while the other uses
a Pentium II 450 MHz processor.
(While the code is the same, the two computers are using different
versions of the compiler and operating system.)
As can be seen from the tables, the implementation of this paper
is faster than the implementation using a hash table for all the lengths
we have considered. For the longest walks on the square lattice it 
is faster by roughly a factor of 80, while on the cubic lattice it 
is faster by roughly a factor of 7 for the longest walks.
The column ``Skip'' gives an idea of how much of the speed-up is a result
of the data structure (i) - (v). For example, table \ref{table_sq_time}
indicates that for the longest walks on the square lattice, the data
structure is responsible for a factor of about 6 in the speed-up.  

Timing computer programs is a tricky business. 
The analysis of the previous section assumes
that we have an unlimited amount of memory, all of which can be 
accessed equally quickly. In a real computer, 
memory effects can cause the actual performance to lag behind the 
theoretical performance. 
The memory needs of our simulations are modest.
Even a million step walk in three dimensions can be stored in RAM, and 
so there is no need to use swap space on the hard drive.
However, most CPU's have a small amount of memory, cache,
that is much  faster than the rest of the RAM. 
For our implementation, 
a large part of the time is spent computing the distance between 
two lattice sites. If the two sites are both stored in the cache,
this computation will be faster than if they are not. 
Thus as $N$ increases, the average time for a single distance computation 
will increase. 
On PC's the cache 
is typically on the order of 256K. We need 8 bytes to store a site on the walk
in two dimensions and 12 bytes in three dimensions. Thus the entire walk 
will fit into a 256K cache for walks up to about 32,000 steps in 
two dimensions and 21,000 steps in three dimensions. 

\begin{figure}[htbp]
  \begin{center}
    \leavevmode
    \epsfxsize=5in 
    \epsfbox{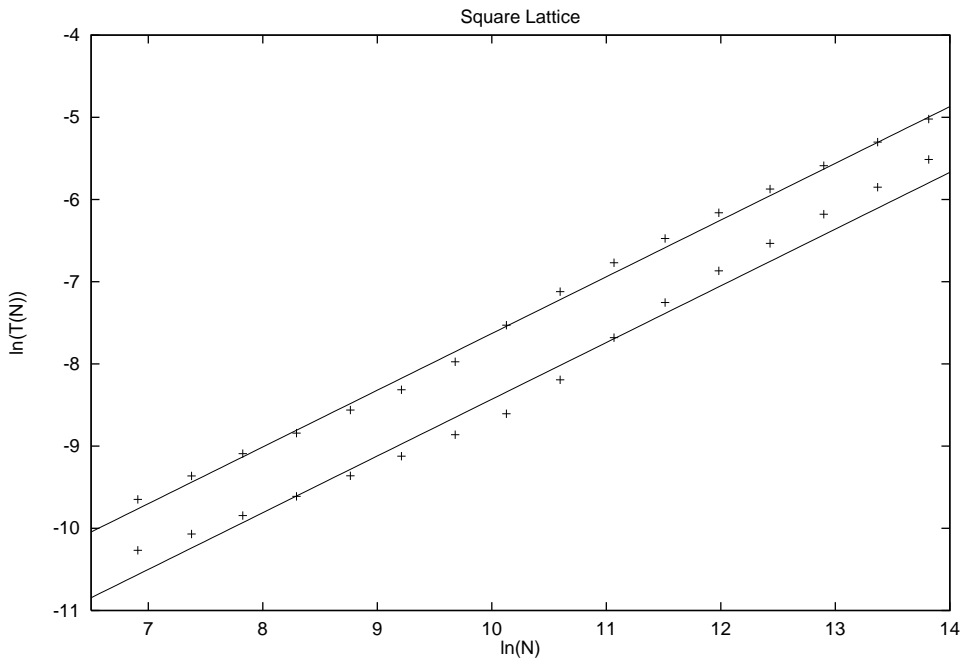}
  \end{center}
  \caption{$T(N)$ is the time required (in secs) 
    per accepted pivot for the square 
    lattice. For the lower sequence of data points, 
    the walk is stored in the usual way. For the upper sequence, the $j$th
    site on the walk is stored in array location $1001*j$ mod 1000001.
    The lines shown have slope 0.69.
   }
  \label{figacceptsq}
\end{figure}

\begin{figure}[htbp]
  \begin{center}
    \leavevmode
    \epsfxsize=5in 
    \epsfbox{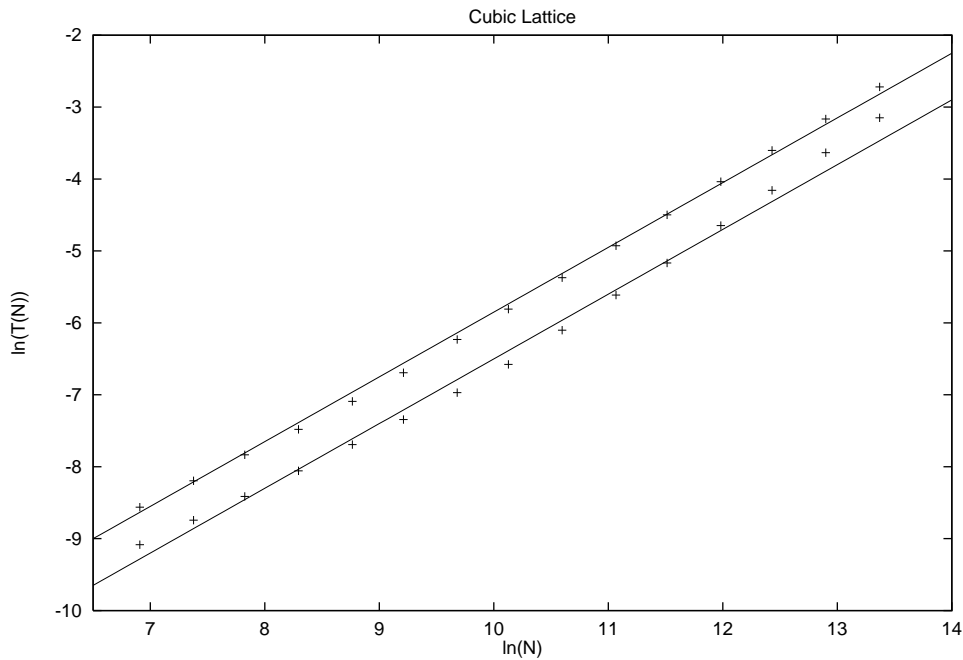}
  \end{center}
  \caption{The time required (in secs) per accepted pivot for the simple cubic
    lattice. As in figure \ref{figacceptsq}, the two curves correspond 
    to two different methods of storing the walk.
    The lines shown have slope 0.91.
   }
  \label{figacceptcubic}
\end{figure}

The lower set of data points in 
figures \ref{figacceptsq} and \ref{figacceptcubic} 
show log-log plots of the time {\it per accepted pivot} as a 
function of $N$ for the square and simple cubic lattices. 
As a guide to the eye, two lines with slope $0.69$ are
shown in figure \ref{figacceptsq} 
and two lines with slope $0.91$ in figure \ref{figacceptcubic}. 
There is a rather abrupt rise in the time required around the 
values of $N$ corresponding to the longest walks that will fit in the 
cache. 

To test if memory effects are playing a significant role, we can force 
the computer to use the same amount of memory for all values of $N$. 
We do this by allocating sufficient memory for 1,000,001 steps,
regardless of the value of $N$. 
We then store $\omega(j)$ in the array location $(1001 j)$ mod $1,000,001$.
This does not 
necessarily make the memory effects uniform for all lengths, but by 
comparing the performance with the original implementation, we can see
if the memory effects are significant. 
The resulting data are the upper sequence of points in 
figures \ref{figacceptsq} and \ref{figacceptcubic}. 
The points are shifted up from the data for 
the implementation in which $\omega(j)$ is simply stored in the $j$th 
array location because every time we access the 
array we must do a multiplication and a mod operation. 
The data for this implementation follows the upper line 
more closely than the previous data followed the lower line.
In comparing the slopes of the lines in the two figures
with our values of $\sigma$,
we should keep in mind that there is an additional factor of 
$\ln(N)$ in the time required coming from the search for $i$ in 
\reff{eqdata}.

We have shown how to implement the pivot algorithm for SAW's so that
the time per accepted pivot grows with the number of steps as $N^q$ with
$q<1$.  
It is difficult to estimate $q$ precisely because of large corrections 
to the $N^q$ behavior. 
For the same reason, the precise value of  
$q$ is irrelevant from a practical point of view. 
Our theoretical analysis and the actual times 
needed to run the algorithm support the conclusion that the 
effective $q$ for values of $N$ that can be simulated is below 1.  

We have restricted our attention to lattice models in this paper, but 
our implementation of the pivot algorithm can be carried out 
for off-lattice (continuum) models as well, provided the length of the 
steps the walk can take is bounded. It would be interesting 
to determine the exponent $\sigma$ in such applications.
In principle, our implementation could also be done for SAW's with 
a nearest neighbor attactive interaction. However, the walks in such 
a model are not as spread out as they are without this interaction.
So the exponent $\sigma$ may be larger for this model.

\bigskip 
\bigskip

\noindent {\bf Acknowledgements:}
The author would like to thank Alan Sokal and the referee 
for many useful comments.
This work was supported by the National Science Foundation (DMS-9970608).

\end{document}